\documentstyle[11pt,graphicx]{article}
\title{  Analytic spectrum of relic gravitational waves
         modified by neutrino free streaming and dark energy}

\author{\small  H.X.  Miao and  Y. Zhang\cite{email} \\
        \small Astrophysics Center \\
       \small University of Science and Technology of China \\
       \small Hefei, Anhui, China }
 \date{}

\begin{document}
\maketitle
\baselineskip=19truept

\def\vek{\vec{k}}
\renewcommand{\arraystretch}{1.5}
\newcommand{\be}{\begin{equation}}
\newcommand{\ee}{\end{equation}}
\newcommand{\ba}{\begin{eqnarray}}
\newcommand{\ea}{\end{eqnarray}}


\sf

\begin{center}
\Large  Abstract
\end{center}

\begin{quote}

We include the effect of neutrino free streaming into the spectrum
of relic gravitational waves (RGWs) in the currently accelerating universe.
For the realistic case of a varying fractional neutrino energy density
and a non-vanishing derivative of mode function at the neutrino decoupling,
the integro-differential equation of RGWs is solved
by a perturbation method for the period from the
neutrino decoupling to the matter-dominant stage.
Incorporating it to the analytic solution of the whole history
of expansion of the universe,
the analytic solution of GRWs is obtained,
evolving from the inflation up to the current acceleration.
The resulting spectrum of GRWs covers the whole range of frequency
$(10^{-19}\sim 10^{10})$Hz,
and improves  the previous results.
It is found that the neutrino free-streaming
causes a reduction of the spectral amplitude
by $\sim 20\%$ in the range $(10^{-16}\sim 10^{-10})$ Hz,
and leaves the other portion of the spectrum almost unchanged.
This agrees with the earlier numerical calculations.
Examination is made on the difference between
the accelerating and non-accelerating models,
and our analysis shows that the ratio of the spectral amplitude
in accelerating $\Lambda$CDM model over that
in  CDM model is $\sim  0.7$,
and within the various  accelerating models
of $\Omega_{\Lambda}> \Omega_m $
the spectral amplitude is proportional to
$ \Omega_m/\Omega_{\Lambda}$ for the whole range of frequency.
Comparison with LIGO S5 Runs Sensitivity
shows that RGWs are not yet detectable by the present LIGO,
and in the future LISA may be able to detect
RGWs in some inflationary models.
\end{quote}


\begin{center}
{\bf 1. Introduction}
\end{center}

Inflationary models generally predict the existence
of a stochastic background of relic gravitational waves (RGWs).
\cite{grishchuk,starobinsky,zhang2}.
Due to their very weak coupling with matter,
RGWs still encode a wealth of information about the very early universe
when they were generated, and enable us to study
the inflationary and the successive physical processes,
much earlier than the recombination time at a temperature $T\sim 0.3$ ev,
up to which CMB information can tell.
Not only the RGWs are the scientific goal of the detections,
such as the laser interferometers now underway
\cite{ligo, lisa,virgo, tamma},
but also are a source, along with the density perturbations,
of CMB anisotropies and polarizations
\cite{basko,Spergel,spergel,seljak,page,hinshaw}.
In particular, the B-polarization of CMB can only be generated by RGWs.
Thus, it is important to calculate the spectrum of RGWs,
which depends  on several physical processes.
First of all, it depends sensitively on the specific
inflationary models \cite{grishchuk,zhang2}.
Moreover, after being generated,
the spectrum of RGWs can be further modified
by the subsequent expansion of the universe,
giving rise to the redshift-suppression on the spectrum.
In our previous analytic and numerical investigations \cite{zhang2},
we studied the RGWs in
the current accelerating expansion of the universe,
obtained the modifications on the spectrum
by the presence of dark energy.
In particular, we have found that
the amplitude of RGWs is reduced by a factor $\sim 0.3$
in comparison with the matter-dominant models,
and that within  the $\Lambda$CDM models
with $\Omega_\Lambda>\Omega_m$,
the amplitude $\propto \Omega_m/\Omega_\Lambda$,
over almost the whole frequency range of the spectrum.

There are other processes that can also change
the spectrum of RGWs.
One important process is the free-streaming of neutrinos that occurred
in the early universe \cite{weinberg}.
It will leave the imprints on the spectrum.
At a temperature $T \sim 2$ Mev during the radiation-dominant stage
in the early universe,
cosmic neutrinos decoupled from electrons and photons,
and started free-streaming in space.
This will give rise to an anisotropic part $\pi_{ij}$
of the energy-momentum tensor $T_{ij}$
as a source of the equation of RGWs,
and will cause a damping effect on the RGWS.
Weinberg analyzed the effect and
arrived at the integro-differential equation for RGWs,
and gave an estimate of the damping on RGWs
due to the neutrinos free-streaming \cite{weinberg}.
Subsequently, in the special case of a constant fractional
neutrino energy density $f_\nu(0)$,
a vanishing time  $u_{dec}=0$ of the neutrino decoupling,
and a vanishing time-derivative $\chi'(u_{dec})=0$,
Dicus and Repko  \cite{dicus} obtained an analytic solution,
in terms of a series of Bessel's functions,
of the integro-differential equation for the radiation stage,
qualitatively agreeing with Weinberg's estimate.
However, this solution holds only for the short wavelength modes
reentering the horizon long after the neutrino decoupling
during radiation-dominant stage,
and the conditions it has used are actually approximations
and will obviously cause some errors.
Moreover, as Weinberg points out,
the solution for the radiation stage is still to be
joined with those for other expansion stages,
so that the effect of neutrino free streaming
is taken into account in a complete computation of the spectrum of RGWs.
In a numerical study of the matter-dominant universe,
Watanabe and Komatsu \cite{yuki} investigated the damping effects
on the  RGWs caused by
the evolution of the effective relativistic degrees of freedom,
including the neutrino free-streaming,
and gave a numerical solution of the
energy density spectrum \cite{yuki}.
But there the important effect of the acceleration
of the present universe has not been considered.

In this paper, extending our previous work on
the analytic spectrum of RGWs
in the accelerating universe \cite{zhang2},
we will include the damping effect of neutrino free-streaming
into our analytic calculation scheme.
Both effects of the accelerating universe and
of the neutrino free streaming are taken into account, simultaneously.
The following improvements are achieved
in this paper over the previous studies.
In comparison with Ref.\cite{yuki},
the damping effect on  the spectrum of RGWs
by the dark energy $\Omega_\Lambda$ is now properly included.
Different from Dicus and Repko's method of series expansion
that is valid in the special case \cite{dicus},
we apply a perturbation method to solve
the integro-differential equation of RGWs by an iterative procedure.
Actually,  for practical use,
the first order solution is enough for an evaluation
for the spectrum of RGWs,
and the solution of higher accuracy can be easily achieved
by going to higher order of iterations.
This calculation has the merit of precisely taking into account of
the time-varying fractional neutrino energy density $f_\nu(u)$,
the non-vanishing time $u_{dec}\neq 0$ and the non-vanishing
time-derivative $\chi'(u_{dec})\neq 0$ of the mode functions.
Therefore, the result is valid for all the modes of RGWs
of an arbitrary wavelength,
and reduces to that in Ref.\cite{dicus} in the special case
for the short wave length limit.
We give the analytic expressions of
the full spectrum $h(k, \eta_H)$ of RGWs itself
and of the spectral energy density $\Omega_g(k)$,
valid for the whole range of frequencies.
As a comprehensive compilation,
by using the parameters $\beta$, $\beta_s$, $\gamma$ and $r$,
respectively,
such important cosmological elements have been explicitly parameterized,
as the inflation, the reheating,
the dark energy, the tensor/scalar ratio.
This will considerably facilitate further studies on the RGWs
and the relevant physical processes.
Besides, several typographical errors in the previous studies
have been corrected thereby.
So not only can it be easily used in computation for
other applications in cosmology,
such as calculations of CMB anisotropies and polarizations
generated by RGWs \cite{page},
but also can be directly compared with
the sensitivity curves of those ongoing and forthcoming
laser interferometer GW detectors,
such as LIGO, LISA, etc \cite{ligo,lisa,virgo, tamma}.

The outline of this paper is as follows.
In section 2, to various stages of expansion of the universe
the scale factor $a(\eta)$ is specified
with the parameters being determined by
the continuity conditions.
In section 3, we present the analytical solutions
of the modes $h_k(\eta)$ of RGWs during each stage,
and, in particular,
include the effect of neutrino free streaming  during
the radiation-dominant stage.
The subtleties of interpreting the observational data
within  the non-accelerating models are discussed.
In section 4, we present  the resulting spectrum and  analyze
the effects of $\beta,\beta_s,\gamma$, $r$
and the neutrino free streaming.
The Appendix gives the detailed calculation of the
anisotropic part $\pi_{ij}$ of the energy-momentum tensor,
and present the perturbation method
for the solution modified by the neutrino free-streaming
during the radiation stage.
In this paper we use unit with $c=\hbar=k_B=1$.

\begin{center}
{\bf 2. Expansion history of the universe}
\end{center}

From the inflationary up to the current accelerating
stage,
the expansion of a spatially flat universe
can be described by the spatially flat
($\Omega_\Lambda +\Omega_m+\Omega_r=1$) Robertson-Walker spacetime
with a metric
\be
ds^2=a^2(\eta)[-d\eta^2+\delta_{ij}dx^idx^j],
\ee
where the scale factor
has the following forms for the successive stages
\cite{grishchuk2000}:

The inflationary stage:
\be \label{inflation}
a(\eta)=l_0|\eta|^{1+\beta},\,\,\,\,-\infty<\eta\leq \eta_1,
\ee
where $1+\beta<0$, and $\eta_1<0$.
The special case of $\beta=-2$ is the
de Sitter expansion of inflation.

The reheating stage:
\be
a(\eta)=a_z|\eta-\eta_p|^{1+\beta_s},\,\,\,\,\eta_1\leq \eta\leq \eta_s,
\ee
here we take the absolute value of $\eta-\eta_p$,
different from Ref.~\cite{zhang2}.
This is because  $1+\beta_s$ might be
negative for some models of the reheating.
As a model parameter, we will mostly take $\beta_s= -0.3$,
though other values are also taken to demonstrate
the effect of the various reheating models.

The radiation-dominant stage:
\be \label{r}
a(\eta)=a_e(\eta-\eta_e),\,\,\,\,\eta_s\leq \eta\leq \eta_2.
\ee
This is the stage during which
the neutrinos decoupled from the radiation component.
We use $\eta_{dec}$ to denote the starting time of
the neutrino decoupling:
$\eta_s<\eta_{dec} < \eta_2$.
The corresponding energy scale is $\sim 2$ Mev
for the decoupling.
As will be seen later,
the wave equation of RGWs is still homogeneous
for $\eta<\eta_{dec}$,
but becomes inhomogeneous for $\eta_{dec} < \eta< \eta_{2}  $.

The matter-dominant stage:
\be \label{m}
a(\eta)=a_m(\eta-\eta_m)^2,\,\,\,\,\eta_2 \leq \eta\leq \eta_E.
\ee

The accelerating stage up to the present time $\eta_H$  \cite{zhang2}:
\be \label{accel}
a(\eta)=l_H|\eta-\eta_a|^{-\gamma},\,\,\,\,\eta_E \leq \eta\leq \eta_H,
\ee
where
the index $\gamma$ depends on the dark energy $\Omega_\Lambda$.
By numerically solving the Friedmann equation \cite{zhang2},
\be \label{Friedmann}
(\frac{a'}{a^2})^2=\frac{8\pi G}{3} (\rho_\Lambda +\rho_m +\rho_r),
\ee
where $a'  \equiv da/d\eta$,
we find that $\gamma\simeq 1.06$ for $\Omega_{\Lambda}=0.65 $,
$\gamma\simeq  1.05$ for $\Omega_{\Lambda}=0.7 $,
and $\gamma\simeq 1.044$ for $\Omega_{\Lambda}=0.75 $
(as a correction to $\gamma\simeq 1.048$ in Ref.\cite{zhang2}).

In the above specifications of $a(\eta)$,
there are five instances of time,
$\eta_1$, $\eta_s$, $\eta_2$, $\eta_E$,
and $\eta_H$,
which separate the different stages.
Four of them  are determined by how much $a(\eta)$ increases
over each stage by the cosmological considerations.
We take the following specifications:
$\zeta_1=\frac{a(\eta_s)}{a(\eta_1)}=300$
for the reheating stage,
$\zeta_s=\frac{a(\eta_2)}{a(\eta_s)}=10^{24}$ for the radiation stage,
$\zeta_2= \frac{a(\eta_E)}{a(\eta_2)}=
\frac{a(\eta_H)}{a(\eta_2)} \frac{a(\eta_E)}{a(\eta_H)}
=3454\zeta_E^{-1}$ for the matter stage,
and
$\zeta_E=\frac{a(\eta_H)}{a(\eta_E)}
=(\frac{\Omega_\Lambda}{\Omega_m})^{1/3}$ for
the present accelerating stage.
Note that here $(\frac{\Omega_\Lambda}{\Omega_m})^{1/3}$
is  model-dependent,
and associated with the value of $\gamma$,
instead of the fixed value (1.33,  as in Ref.\cite{zhang2}).
The remaining time instance is fixed by an overall normalization,
namely
\be \label{norm}
|\eta_H-\eta_a|=1.
\ee
There are twelve  constants in the expressions of $a(\eta)$,
among which  $\beta$, $\beta_s$ and $\gamma$
are imposed as the model parameters,
for the inflation, the reheating, and the acceleration,
respectively.
So there remain nine  constants.
By the continuity of $a(\eta)$ and of $a(\eta)'$
at the four given joining
points $\eta_1$, $\eta_s$, $\eta_2$ and $\eta_E$,
one can fix eight constants.
Only one constant remains,
which can be fixed by the present expansion rate $H_0$
of the universe and Eq.(\ref{norm}),
\be
l_H=\gamma/H_0.
\ee
Then all parameters are fixed as the following:
\begin{eqnarray} \label{aling1}
&&\eta_a-\eta_E=\zeta_E^{\frac{1}{\gamma}},\nonumber\\
&&\eta_E -\eta_m=\frac{2}{\gamma}
                \zeta_E^{\frac{1}{\gamma}},\nonumber\\
&&\eta_2-\eta_m=\frac{2}{\gamma}
      \zeta_2^{-\frac{1}{2}}\zeta_E^{\frac{1}{\gamma}},\nonumber\\
&&\eta_2-\eta_e=\frac{1}{\gamma}\zeta_2^{-\frac{1}{2}}
      \zeta_E^{\frac{1}{\gamma}},              \nonumber\\
&&\eta_s-\eta_e=\frac{1}{\gamma}\zeta_s^{-1}
                 \zeta_2^{-\frac{1}{2}}
                 \zeta_E^{\frac{1}{\gamma}},\nonumber\\
&&\eta_s-\eta_p=\frac{1}{\gamma}(1+\beta_s)
                 \zeta_s^{-1}\zeta_2^{-\frac{1}{2}}
                  \zeta_E^{\frac{1}{\gamma}},\nonumber\\
&&\eta_1-\eta_p=\frac{1}{\gamma}(1+\beta_s)
      \zeta_1^{\frac{-1}{1+\beta_s}}\zeta_s^{-1}
      \zeta_2^{-\frac{1}{2}}\zeta_E^{\frac{1}{\gamma}}, \nonumber\\
&&\eta_1=\frac{1}{\gamma}(1+\beta)
      \zeta_1^{\frac{-1}{1+\beta_s}}\zeta_s^{-1}
      \zeta_2^{-\frac{1}{2}}\zeta_E^{\frac{1}{\gamma}},\nonumber\\
&&\eta_{dec}=1.15\times10^{-10}\zeta_E\zeta_2\eta_2,
\end{eqnarray}
and
\begin{eqnarray} \label{aling2}
&&a_m=\frac{l_H}{4}\,\gamma^2\,\zeta_E^{-(1+\frac{2}{\gamma})},\nonumber\\
&&a_e=l_H\,\gamma\,\zeta_2^{-\frac{1}{2}}
    \zeta_E^{-(1+\frac{1}{\gamma})},\nonumber\\
&&a_z=l_H\,\gamma^{1+\beta_s}|1+\beta_s|^{-(1+\beta_s)}
    \zeta_s^{\beta_s}\zeta_2^{\frac{\beta_s-1}{2}}
    \zeta_E^{-(1+\frac{1+\beta_s}{\gamma})},\nonumber\\
&&l_0=l_H\,\gamma^{1+\beta}\,|1+\beta|^{-(1+\beta)}
    \zeta_1^{\frac{\beta-\beta_s}{1+\beta_s}}
    \zeta_s^{\beta}\zeta_2^{\frac{\beta-1}{2}}
    \zeta_E^{-(1+\frac{1+\beta}{\gamma})}.
\end{eqnarray}
The above expressions
correct some typographical errors in Ref.\cite{zhang2}.

In the expanding universe, the physical wavelength is related to
the comoving wavenumber $k$ by
\be
\lambda\equiv \frac{2\pi a(\eta)}{k},
\ee
and the wavenumber $k_H$  corresponding to
the present Hubble radius is
\be
k_H = \frac{2\pi a(\eta_H)}{1/H_0}=2\pi \gamma.
\ee
There is another wavenumber
\be
k_E \equiv \frac{2\pi a(\eta_E)}{1/H_0}=\frac{k_H}{1+z_E},
\ee
whose corresponding wavelength at the time $\eta_E$
is the Hubble radius $1/H_0$.
In the present universe
the physical frequency corresponding to a wavenumber $k$
is given by
\be \label{12}
\nu = \frac{1}{\lambda}= \frac{k}{2\pi a (\eta_H)}
= \frac{H_0}{2\pi\gamma} k.
\ee

\begin{center}
{\bf 3. Analytical solution}
\end{center}

In the presence of
the gravitational waves,
the perturbed metric  is
\be
ds^2=a^2(\eta)[-d\eta^2+(\delta_{ij}+h_{ij})dx^idx^j],
\ee
where the tensorial perturbation $h_{ij}$
is a $3\times 3$ matrix and is taken to
be transverse and traceless
\begin{eqnarray}
h^i_{\,\,i}=0,&&h_{ij,j}=0.
\end{eqnarray}
The wave equation of RGWs is
\be \label{eq1}
 \partial_{\nu}( \sqrt{-g}\partial^{\nu} h_{ij} )=0.
\ee
However,
 from the temperature $T \simeq 2 $ Mev up to the beginning
of the matter domination,
the neutrinos are decoupled from electrons and photons and
start to freely stream in space.
This effect of neutrino free streaming
gives rise to an anisotropic portion $\pi_{ij}$
of the energy-momentum stress $T_{ij}$.
Then Eq.(\ref{eq1}) acquires
an inhomogeneous source term $-16\pi G\pi_{ij}$ on the right hand side
during the period $\eta_{dec}<\eta< \eta_2$.
As is shown in the Appendix,
the anisotropic stress $\pi_{ij}$ is also transverse and
traceless, and it is zero before the decoupling
and becomes negligible small
after the matter domination.
To solve the equation, we decompose $h_{ij}$  into the Fourier
modes of the comoving wave number $k$ and into the polarization
state $\sigma$ as
\be
\label{planwave}
h_{ij}(\eta,{\bf x})=
\sum_{\sigma}\int\frac{d^3k}{(2\pi)^3}
\epsilon^{\sigma}_{ij}h_k^{(\sigma)}(\eta)
e^{i\bf{k}\cdot{x}}
\, ,
\ee
where
$h_{-k}^{(\sigma)*}(\eta)=h_k^{(\sigma)}(\eta)$
ensuring that $h_{ij}$ be real,
$\epsilon^{\sigma}_{ij}$ is the polarization tensor,
 and $\sigma$ denotes the polarization states $\times,+$.
Here $h_{ij}$ is  treated as a classical field,
instead of a quantum operator \cite{grishchuk, zhang2}.
In terms of the mode $h^{(\sigma)}_{k}$,
Eq.(\ref{planwave}) reduces to
\be \label{eq}
h^{ (\sigma) }_{k}{''}(\eta)
+2\frac{a'(\eta)}{a(\eta)}h^{ (\sigma) }_k {'}(\eta)
+k^2h^{(\sigma)}_k(\eta)=0.
\ee
Since for each polarization, $\times$,  $+$,
the wave equation is the same
and has the same statistical properties,
from now on the super index $(\sigma)$ can be dropped from
$h^{(\sigma)}_k$.
As demonstrated in Eq.(\ref{inflation}) through Eq.(\ref{accel}),
for all the stages of expansion
the time-dependent scale factor
is of a generic form
\be
a(\eta) \propto \eta^\alpha,
\ee
the solution to Eq.(\ref{eq})
is a linear combination of
Bessel function $J_{\nu}$
and Neumann function $N_{\nu}$
\be \label{hom}
h_k(\eta)=x^{\frac{1}{2}-\alpha}
 \big[a_1 J_{\alpha-\frac{1}{2}}(k \eta)
+a_2   N_{\alpha-\frac{1}{2}}(k \eta)\big],
\ee
where the constants $a_1$ and $a_2$ are determined
by the continuity of $h_k$ and of $h'_k$
at the joining points
$\eta_1,\eta_s,\eta_2$ and $\eta_E$.
However, as mentioned earlier, during the neutrino free streaming
with $\eta_{dec} \leq \eta\le\eta_{2}$,
Eq.(\ref{eq}) will be modified and
its solution will  be  given later.

The inflationary stage  has the solution
\be \label{infl}
h_k(\eta)=A_0 l_0^{-1}|\eta|^{-\frac{1}{2}-\beta}
\big[ A_1 J_{\frac{1}{2}+\beta}(x)
     +A_2 J_{-(\frac{1}{2}+\beta)}(x) \big],
\,\,\,\, -\infty<\eta\leq \eta_1
\ee
where $x\equiv k\eta$ and
\be
A_1=-\frac{i}{\cos \beta\pi}\sqrt{\frac{\pi}{2}}e^{i\pi\beta/2},
\,\,\,\,\,
A_2=iA_1e^{-i\pi\beta},
\ee
are taken \cite{grishchuk1993},
so that the so-called \textit{adiabatic vacuum} is achieved:
$\lim_{k\rightarrow \infty}h_k(\eta)\propto e^{-ik\eta} $
in the high frequency limit \cite{parker}.
Moreover, the constant $A_0$ in Eq.(\ref{infl}) is
independent of $k$, whose value  is determined by
the initial amplitude of the spectrum,
so that for $k\eta\ll 1$ the $k$-dependence of $h_k(\eta)$ is given by
\be
h_k(\eta)\propto J_{\frac{1}{2}+\beta}(x)\propto
 k^{\frac{1}{2}+\beta}.
\ee
As will be seen,
this choice will lead to the required initial spectrum
in Eq.(\ref{initialspectrum}).

The reheating stage  has
\be \label{reh}
h_k(\eta)=t^{-\frac{1}{2}-\beta_s}
\Big[B_1J_{\frac{1}{2}+\beta_s}(k\,t)
+B_2N_{\frac{1}{2}+\beta_s}(k\,t)\Big],
\,\,\,\,\,  \eta_1\leq  \eta\leq\eta_s
\ee
where  $t\equiv \eta-\eta_p$ and
\begin{eqnarray}
&&B_1=\frac{-1}{2}\pi\,t_1^{\frac{3}{2}+\beta_s}
\big[kN_{\frac{3}{2}+\beta_s}(k\,t_1)h_k(\eta_1)+
N_{\frac{1}{2}+\beta_s}(k\,t_1)h_k'(\eta_1)\big],\\
&&B_2=\frac{1}{2}\pi\,t_1^{\frac{3}{2}+\beta_s}
\big[kJ_{\frac{3}{2}+\beta_s}(k\,t_1)h_k(\eta_1)+
J_{\frac{1}{2}+\beta_s}(k\,t_1)h_k'(\eta_1)\big],
\end{eqnarray}
with $t_1\equiv \eta_1-\eta_p$,
and  $h_k(\eta_1 )$ and  $h_k'(\eta_1)$
are the corresponding values from the precedent inflation stage.

The radiation-dominant stage needs to be divided into
two parts.
The first part of the stage is before the neutrino decoupling
when $\eta_s\leq \eta\leq\eta_{dec}$,
the neutrino damping is ineffective yet,
the wave equation is still homogenous
with the solution
\begin{equation} \label{r1}
h_k(\eta)=y^{-\frac{1}{2}}
\Big[C_1J_{\frac{1}{2}}(k\,y)+C_2N_{\frac{1}{2}}(k\,y)\Big],
\,\,\,\,\, \eta_s \leq \eta\leq\eta_{dec}
\end{equation}
where   $y\equiv \eta-\eta_e$ and
\begin{eqnarray}
&&C_1=\frac{-1}{2}\pi\,y_s^{\frac{3}{2}}
\big[kN_{\frac{3}{2}}(k\,y_s)h_k(\eta_s)+
N_{\frac{1}{2}}(k\,y_s)h_k'(\eta_s)\big],\\
&&C_2=\frac{1}{2}\pi\,y_s^{\frac{3}{2}}
\big[kJ_{\frac{3}{2}}(k\,y_s)h_k(\eta_s)+
J_{\frac{1}{2}}(k\,y_s)h_k'(\eta_s)\big],
\end{eqnarray}
where  $y_s\equiv \eta_s-\eta_e$,
and $h_k(\eta_s )$ and  $h_k'(\eta_s)$ are
from the reheating stage.
If we do not include the neutrino effect and
let Eq.(\ref{r1}) be valid for the whole radiation stage
$\eta_s\leq  \eta \leq\eta_2$,
then our previous exact result  \cite{zhang2}
would be recovered.

\begin{figure}
\centerline{\includegraphics[width=8cm]{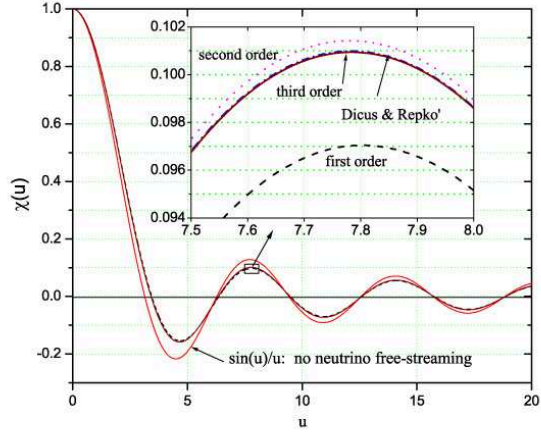}}
\caption{\label{app vs exact2}
Under the approximations in Eq.(\ref{apprxmdicus}),
the solutions by our method are compared with that
in Ref.\cite{dicus}. }
\end{figure}

The second part is from the neutrino decoupling
up to the matter domination with $\eta_{dec}\leq \eta\leq\eta_2$.
The temperature at the neutrino decoupling $\eta_{dec}$
is taken to be $T\simeq 2$ Mev.
During this period the wave equation is
\be \label{imheq}
h_{k}''(\eta)
+2\frac{a'(\eta)}{a(\eta)}h_k'(\eta)
+k^2h_k(\eta)=16\pi Ga^2\pi_k(\eta) .
\ee
The detailed derivation  are given in the Appendix \cite{weinberg}.
Writing  the  mode function as
\be \label{h}
h_k(\eta) =
h_k(\eta_{dec})\chi(\eta), \,\,\,\,\, \eta_{dec}\leq \eta\leq\eta_2
\ee
where $h_k(\eta_{dec})$ is given by Eq.(\ref{r1})
evaluated at $\eta_{dec}$,
and $\chi(u)$ satisfies
the following integro-differential equation
\be\label{eq2}
\chi''(u)+\frac{2}{u}\chi'(u)+\chi(u)=-24
\frac{f_\nu (0)}{u^2(1+\alpha\,u)}
\int_{u_{dec}}^{u}d U K(u-U)\chi'(U),
\ee
where $u\equiv k\eta$,
$f_{\nu}(0)=0.40523$ is the fractional energy density of
neutrinos at $u=0$,
$\alpha\equiv a_e/(k\,a(\eta_2))
=k^{-1}\gamma \zeta^{1/2}_2\zeta_E^{1/\gamma}$,
and $K(u)$ is the kernel defined in Eq.(\ref{kernel}) in the Appendix.
In dealing with Eq.(\ref{eq2}) Dicus \& Repko \cite{dicus}
use the following approximations
\be \label{apprxmdicus}
\alpha=0, \,\,\,\, u_{dec}=0, \,\,\,\,   \chi'(\eta_{dec})=0,
\ee
and derive  an analytical solution,
valid for those modes reentering the horizon
long after the neutrino decoupling during
the radiation-dominant stage.
Here without making the approximations in Eq.(\ref{apprxmdicus}),
we try to give a solution valid for all the modes
that reenter the horizon both before and after
the neutrino decoupling.
The idea is that,  the neutrino damping effect is small,
therefore the right hand side of Eq.(\ref{eq2})
will cause only a small variation to
the  homogeneous solution $\chi_0(u)$.
Thus,  as an approximation,
one substitutes $\chi_0(u)$ in place of $\chi(u)$
in the integration on the right hand side of Eq.(\ref{eq2}),
and obtains the first order approximate solution.
As it turns out, this is accurate enough for our purpose
of calculating the spectrum.
For the higher order solutions
this process can be  iterated to achieve higher accuracy.
The Appendix gives the detailed expressions
of the solutions.
To examine this perturbation method,
Figure \ref{app vs exact2} plots  the mode function $\chi(u)$
as the solution of different orders,
respectively, where the same assumption as Eq.(\ref{apprxmdicus})
are adopted to compare
with Dicus \& Repko's  analytic result \cite{dicus}.
It is seen that the first, and second order solutions
differ by $\sim 4\%$,
and by $\sim 1\%$, respectively, from the exact one,
the third order solution almost overlaps with it.
Therefore, our method is effective in evaluating the damping
caused by neutrino free-streaming.
Moreover, our result, Eqs.(\ref{first}) and (\ref{second})
in the Appendix,
holds for any wavelength and
for the realistic condition of
$\alpha \neq 0$, $u_{dec}\neq 0$, and $\chi'(\eta_{dec}) \neq0$,
while the approximation in Refs.\cite{weinberg} and \cite{dicus}
is not valid for the very short nor for the long modes.
Figure  \ref{app vs exact2} shows that
the solution without the neutrino free-streaming
has a higher amplitude, as expected.

\begin{figure}
\centerline{\includegraphics[width=8cm]{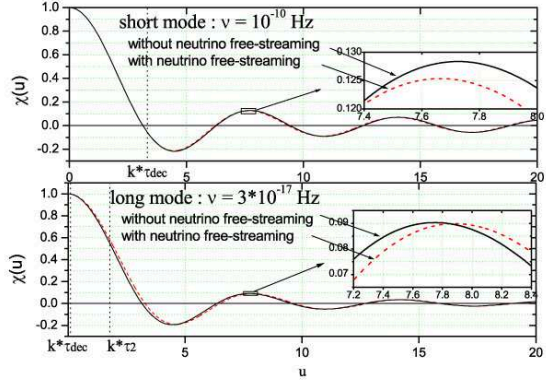}}
\caption{\label{longshort}
Neutrino free-streaming   barely affects
the  short (upper) and the long modes  (lower). }
\end{figure}

Our calculation reveals that
the neutrino damping on the RGWs is mainly pronouncing only
in the  frequency range $\nu\simeq ( 10^{-16}\sim 10^{-10})$ Hz,
which corresponds to $ k \simeq (10^2\sim  10^8)$ and
 $ \alpha \simeq ( 10^{-7} \sim  10^{-1})$.
Outside this range the neutrino damping barely alters the RGWs.
Figure \ref{longshort} shows that our solution of
 short ($\nu\geq 10^{-10}$Hz) and long ($\nu \leq 10^{-16}$ Hz) modes
almost overlap the homogeneous solution
without neutrino free-streaming.
For the short modes reentering the horizon well before the decoupling,
the factor $1/u^2 $ on the r.h.s. of Eq.(\ref{77})
is very small and the inhomogeneous term is negligible.
For those long modes,
they are still outside the horizon during the neutrino free streaming,
and are not affected by the damping.
Only much later do these modes reenter the horizon,
the neutrino density $f_{\nu}(u)$ becomes negligibly small,
the homogeneous solution is valid for these long  modes.
Therefore, in long and short wavelength limit, the solution for
RGWs is practically that of the homogeneous equation.
Moreover, the upper panel of Figure \ref{longshort} shows that,
at the neutrino decoupling time $\eta_{dec}$,
the time derivative of the short mode function
 $\chi'(\eta_{dec})$ deviates from zero considerably.
Therefore, the approximation in Eq.(\ref{apprxmdicus})
is not accurate enough for the short modes of RGWs.


The matter-dominant  stage has
\begin{equation}  \label{matter}
h_k(\eta)=z^{-\frac{3}{2}}
\Big[D_1J_{\frac{3}{2}}(k\,z)+D_2N_{\frac{3}{2}}(k\,z)\Big],
\,\,\,\, \eta_2\leq \eta\leq\eta_E
\end{equation}
where $z\equiv \eta-\eta_m$ and
\begin{eqnarray}
&&D_1=\frac{-1}{2}\pi\,z_2^{\frac{5}{2}}
\big[kN_{\frac{5}{2}}(k\,z_2)h_k(\eta_2)+
N_{\frac{3}{2}}(k\,z_2)h_k'(\eta_2)\big],\\
&&D_2=\frac{1}{2}\pi\,z_2^{\frac{5}{2}}
\big[kJ_{\frac{5}{2}}(k\,z_2)h_k(\eta_2)+
J_{\frac{3}{2}}(k\,z_2)h_k'(\eta_2)\big],
\end{eqnarray}
withe $z_2\equiv \eta_2-\eta_m$.
In the expressions of $D_1$ and $D_2$,
the mode functions $h_k(\eta_2)$ and $h'_k(\eta_2)$ are again
from the precedent stage.

The accelerating stage  has
\be \label{acc}
h_k(\eta)=s^{\frac{1}{2}+\gamma}
\Big[E_1J_{-\frac{1}{2}-\gamma}(k\,s)
+E_2N_{-\frac{1}{2}-\gamma}(k\,s)\Big],
\,\,\,\,\, \eta_E\leq\eta\leq\eta_H
\ee
where $s\equiv \eta-\eta_a$  and
\begin{eqnarray}
&&E_1=\frac{-1}{2}\pi\,s_E^{\frac{1}{2}-\gamma}
\big[kN_{\frac{1}{2}-\gamma}(k\,s_E)h_k(\eta_E)+
N_{-\frac{1}{2}-\gamma}(k\,s_E)h_k'(\eta_E)\big],\\
&&E_2=\frac{1}{2}\pi\,s_E^{\frac{1}{2}-\gamma}
\big[kJ_{\frac{1}{2}-\gamma}(k\,s_E)h_k(\eta_E)+
J_{-\frac{1}{2}-\gamma}(k\,s_E)h_k'(\eta_E)\big],
\end{eqnarray}
with  $s_E\equiv \eta_E-\eta_a$.
So far, the explicit solution of $h_k(\eta)$ has been obtained
for all the  expansion stages,
from Eq.(\ref{infl}) through Eq.(\ref{acc}).

\begin{figure}
\centerline{\includegraphics[width=8cm]{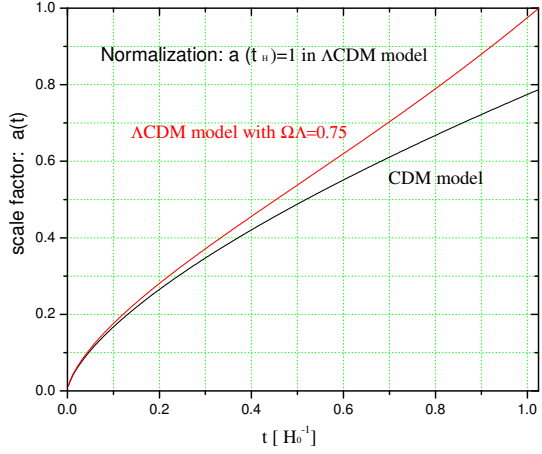}}
\caption{\label{scale factor}
The scale factor $a(t)$ in
the accelerating,  and non-accelerating models, respectively. }
\end{figure}

The above detailed expressions of $h_k(\eta)$
are the major ingredients to determine
the the spectrum of RGWs in the accelerating universe.
What kind of RGWS  would a matter-dominant universe have?
To compare with the spatially flat accelerating universe,
this non-accelerating universe is assumed to be also spatially flat
with $\Omega_m+\Omega_r =1$.
It should also go through the consecutive expansion
stages listed previously,
from the inflation to the matter-dominant,
except the  accelerating stage that is replaced by a continuation of
the matter-dominant stage up to the present time  $\eta_H$.
In each stage the mode function $h_k(\eta)$ is of
the same form as those given in
Eqs.(\ref{infl}), (\ref{reh}), (\ref{r1}),
(\ref{h}), and (\ref{matter}), respectively.
But the time duration of the matter stage
for Eq.(\ref{matter}) is now extended to $\eta_2\leq\eta\leq\eta_H$.
In both the accelerating and
the matter-dominant models the mode $h_k(\eta)$ is sensitive to
the scale factor $a(\eta)$ determined by
their respective Friedmann equation (\ref{Friedmann}),
in which one sets $\rho_\Lambda =0$ for the matter-dominant model.
To have a specific comparison of the two models,
let us start at the time $\eta_2$ of the equality of
radiation-matter with $1+z=3454$,
when $\rho_\Lambda \ll \rho_m=\rho_r$,
and  it can be assumed that both
models have the same initial values $a(\eta_2)$ and $a'(\eta_2)$.
Instead of the sudden transition approximation
as in Eqs.(\ref{m}) and (\ref{accel}),
we solve numerically the Friedmann equation
in both models  up to the present time $\eta_H$ with $z=0$.
Doing this is equivalent to assuming that both models would have
an equal  age of the universe.
As a result, it is found that the scale factor $a(\eta_H)$
in the accelerating model is $\sim 1.3$ times of
that in the matter-dominant model shown in Fig.\ref{scale factor}.
This difference of $a(\eta_H)$ will consequently cause
a difference in the spectra for the two models.
As is  known \cite{grishchuk, zhang2},  inside the horizon
the amplitude of modes $h_k(\eta)\propto 1/a(\eta)$,
so the matter-dominant model would predict a spectral amplitude
higher than the accelerating model.
Indeed,  our analytic calculation demonstrates that the ratio
of the spectral amplitudes of CDM over those of
$\Lambda$CDM is $\sim  1.3$.
Moreover, we like to emphasize that
there are some subtleties with the matter-dominant models,
regarding to interpretation of the current cosmological observations.
The actual universe is an accelerating one,
so the observed  Hubble constant is properly interpreted as
the current expansion rate  in the accelerating model,
$H_0=a'/a^2 (\eta_H)$.
However, as our calculation has shown,
the virtual matter-dominant  universe would
have  a smaller rate $a'/a^2(\eta_H) \simeq 0.65\, H_0$.
In this regard, Ref.\cite{yuki} uses the observed Hubble constant $H_0$
as the current expansion rate of the virtual matter-dominant
universe.
This  would give a spectrum with amplitude lower
by an extra factor $\sim 1.3$ than it should have.

\begin{center}
 {\bf 4. Spectrum of relic gravitational waves}
\end{center}

The spectrum of RGWs $h(k,\eta)$ at a time $\eta$
is defined by the following equation \cite{grishchuk2000}:
\be
\int_0^{\infty}h^2(k,\eta)\frac{dk}{k}\equiv\langle0|
h^{ij}(\textbf{x},\eta)h_{ij}(\textbf{x},\eta)|0\rangle,
\ee
where the right-hand side is the expectation value of the
$h^{ij}h_{ij}$.
Calculation  yields the spectrum, which is related
to the mode function $h_k(\eta)$ as follows
\be \label{relation}
h(k,\eta)=\frac{2}{\pi}k^{3/2} |h_k(\eta)|,
\ee
where the factor $2$ counts for the two independent polarizations.
At present with time $\eta_H$ the spectrum is
\be \label{33}
h(k,\eta_H)  = \frac{2}{\pi}k^{3/2}|h_k(\eta_H)|.
\ee
Note that this expression is formally different from
the previous one in Refs.\cite{zhang2}
only because here we use a different expansion
for $h_{ij}(\eta,{\bf x})$ in Eq.(\ref{planwave}).
One of the most important properties
of the inflation is that
the initial spectrum of GRWs at the time $\eta_i$ of
the horizon-crossing
during the inflation is nearly scale-invariant \cite{grishchuk2000}:
\be \label{initialspectrum}
h(k,\eta_i) =A(\frac{k}{k_H})^{2+\beta},
\ee
where $2+\beta\simeq 0$,
and $A$ is a $k$-independent constant
to be fixed by the observed CMB anisotropies  in practice.
The First Year WMAP gives
the scalar spectral index $n_s=0.99\pm 0.04$ \cite{Spergel}.
The Three Year WMAP gives    $n_s=0.951^{+0.015}_{-0.019}$ \cite{spergel},
while in combination with constraints from  SDSS, SNIa,
and the galaxy clustering,
it would give $n_s=0.965\pm 0.012$ (68\%  CL) \cite{seljak}.
From  the relation $n_s=2\beta+5$ \cite{grishchuk, zhang2},
we have the inflation index $\beta=-2.02$ for $n_s=0.951$.
Note that the constant $A$
is directly proportional to  $A_0$ in Eq.(\ref{infl})
through the relation (\ref{relation}).
Since the observed CMB anisotropies \cite{Spergel}
is $\Delta T/T \simeq 0.37\times 10^{-5}$ at $l\sim2$,
which corresponds to anisotropies on scales of the Hubble radius $1/H_0$,
so,  as in Refs.\cite{zhang2},
we take the  normalization of the spectrum
\be \label{norml}
h(k_E,\eta_H)=0.37\times10^{-5}r^{\frac{1}{2}},
\ee
where  $k_E =\frac{k_H}{(1+z_E)}= \frac{2\pi \gamma}{(1+z_E)}$
is the wave number that crosses the horizon at $\eta_E$,
its corresponding  physical frequency  being
$\nu_E = k_E/2\pi a(\eta_H)=H_0/(1+z_E)\sim 1 \times 10^{-18}$ Hz,
$r$ is taken as a parameter roughly representing
the tensor/scalar ratio.
In Eq.(\ref{norml}) it is $r^{\frac{1}{2}}$
rather than $r$ as in Ref.\cite{zhang2}.
The value of the ratio $r$ is an important issue
and is still unsettled yet.
However, as examined in details in Ref.\cite{baskaran},
the relative contributions from the RGWs and from
the density perturbations are, in fact,
frequency-dependent;
thus, generally speaking,
for different frequency ranges $r$ can take on
a different values.
Therefore, in our treatment, for simplicity,
$r$ is only taken as a constant parameter for normalization of RGWs,
and does not accurately represent the actual relative contributions.
Currently, only observational constraints on  $r$ have been given.
Recently the Three Year WMAP constraint is $r<2.2 $ ($95\%$ CL)
evaluated at $k=0.002$ Mpc$^{-1}$, and
the full WMAP constraint is $r<0.55 $ (95\% CL) \cite{page}.
The combination from such observations,
as of the Lyman-$\alpha $ forest
power spectrum from SDSS, 3-year WMAP,  supernovae SN,
and galaxy clustering,
gives an upper limit $r<0.22$ ($95\%$ CL) \cite{seljak}.
Moreover, the ratio $r$ may be allowed to take on
different values on different range of frequency,
but we will take a constant $r$ for simplicity.

The spectral energy density $\Omega_g(k)$
of the RGWs is given by
\be\label{32}
\Omega_g(k)=\frac{\pi^2}{3}h^2(k,\eta_H)\Big(\frac{k}{k_H}\Big)^2,
\ee
directly associated with the spectrum of RGWs
$h(k,\eta_H)$ in Eq.(\ref{33}).
This follows from the definition  \cite{grishchuk2000, baskaran}
\be \label{gwe}
\Omega_{GW} \equiv \frac{\rho_g}{\rho_c} =
\int_{k_{low}}^{k_{upper}} \Omega_g(k)\frac{dk}{k},
\ee
where $\rho_g=\frac{1}{32\pi G}h_{ij,0}h^{ij}_{,0}$
is the energy density of RGWs,
and $\rho_c=3H_0^2/8\pi G$ is the critical energy density.
The integration in Eq.(\ref{gwe})
has the lower and upper limits, $k_{low}$ and  $k_{upper}$,
as the cutoffs of the wavenumber.
For the lower limit $k_{low}$,
the corresponding wavelength may be taken to be
the current Hubble radius,
$\lambda_{low}=1/H_0$.
This is because the waves with wavelengths longer than $1/H_0$
should be treated as part of the space-time background
and should not be included to
the energy of RGWS \cite{grishchuk} \cite{zeldovich}.
By Eq.(\ref{12}), the corresponding frequency
\be \label{lowerlimit}
\nu_{low} \simeq 2\times 10^{-18} {\, \rm Hz}.
\ee
The upper limit $k_{upper}$ can be determined by
as the following.
During the inflation the modes of GRWs with wavenumbers greater
than the expansion rate $H(\eta_i)$
are  approaching the adiabatic limit, therefore,
their generation is thus effectively suppressed \cite{parker}.
Taking the scale of the vacuum energy driving the inflation to
be  $E_{vac}\sim 10^{16}$ Gev, typical of Grand Unified Theories,
then $H(\eta_i)\sim 10^{13}$ Gev $\simeq 10^{38}$ Hz.
During the subsequent stages of cosmic expansion,
the corresponding frequency $\nu$ of this value
will be redshifted by a factor $a(\eta_i)/a(\eta_H)\sim 10^{-29}$;
thus one has
\be\label{upperlimit}
\nu_{upper} \simeq 10^{10} {\, \rm Hz}.
\ee
If the energy scale for the inflation is lower than $10^{16}$ Gev,
then the upper limit $\nu_{upper}$ will
be lower than that in Eq.(\ref{upperlimit}) correspondingly.
These lower and upper  integration  limits
in Eqs.(\ref{lowerlimit}) and (\ref{upperlimit})
also ensure the convergence of the integration of Eq.(\ref{gwe}).

In the absence of direct detection of RGWs,
the constraints on the energy density $\Omega_{GW}$
is more relevant.
Given the model parameters $\beta$, $\beta_s$, $ \gamma$,
and $r$, the definite integration of Eq.(\ref{gwe}) yields
 $\Omega_{GW}$ of RGWs.
For the fixed parameters $r=0.22$, $\Omega_\Lambda =0.75$,
and $\beta_s=-0.3$,
one finds  $\Omega_{GW}=1.12\times10^{-2}$ for the inflationary
model of $\beta=-1.8$.
Such a large energy density will inevitably affect the
expansion rate of the universe at a temperature $T\sim$
a few MeV when the nucleosynthesis process is going on.
The nucleosythesis bound is
\cite{maggiore}
\be \label{BBN}
\Omega_{GW}  h^2 <8.9\times 10^{-6}
\ee
with $h\sim 0.71$ being the Hubble parameter \cite{Spergel}.
Thus the $\beta=-1.8$ model
with $r=0.22$ predicts an energy density
$\Omega_{GW}$ being some four orders higher than the upper bound
given Eq.(\ref{BBN}).
So this model will be in jeopardy,
unless the parameter $r$ is much smaller than $0.22$.
Under the same set of parameters $r=0.22$, $\Omega_\Lambda =0.75$,
and $\beta_s=-0.3$,
the  model of $\beta=-1.9$ gives  $\Omega_{GW}=2.04\times10^{-8}$,
and the model of  $\beta=-2.02$ gives $\Omega_{GW}= 1.54\times10^{-14}$,
both models are  safely below the  nucleosythesis bound in Eq.(\ref{BBN}).

In the following we demonstrate the details of
the resulting spectra $h(k,\eta_H)$ and $\Omega_g(k)$
of RGWs,
their explicit dependence upon
the model parameters $\beta$, $\beta_s$, $ \gamma$,
and the modifications by the neutrino damping.

\begin{figure}
\centerline{\includegraphics[width=8cm]{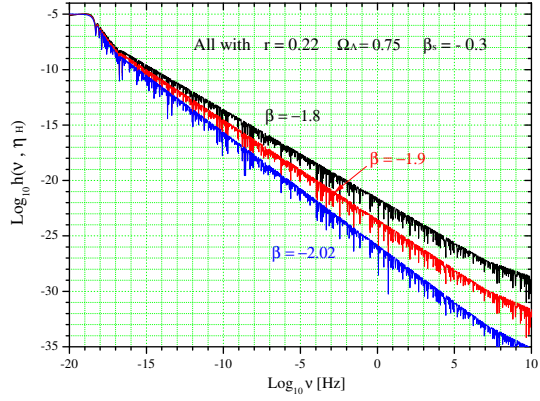}}
\caption{\label{diff_beta}
The spectrum $h(\nu, \eta_H)$ of GRW is very sensitive to
the inflation parameter $\beta$. }
\end{figure}

Figure \ref{diff_beta} gives the spectrum $h(\nu, \eta_H)$
as a function of the frequency $\nu$
without neutrino free streaming.
To show the dependence upon the inflationary models,
for the fixed $r=0.22$, $\Omega_\Lambda=0.75$, $\beta_s=-0.3$,
we plot $h(\nu,\eta_H)$ in three models of
$\beta = -1.8$, $-1.9$, and $-2.02$.
It is seen that $h(\nu, \eta_H)$ is very sensitive to $\beta$.
A smaller $\beta$ will generate
less power of RGWs for all frequencies.
The details of the spectrum is similar to that given
in Refs.\cite{zhang2}.

\begin{figure}
\centerline{\includegraphics[width=8cm]{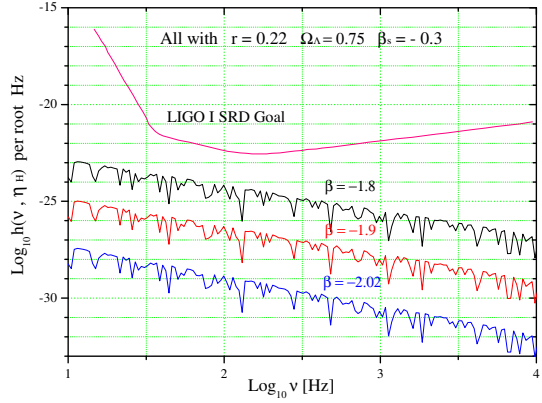}}
\caption{\label{ligo_sensitivity}
 Comparison of the spectra with
the LIGO I SRD Goal sensitivity curve that has already been
achieved by S5 of LIGO~\cite{ligo}.
The vertical axis is the r.m.s amplitude per root Hz
defined in Eq.(\ref{rmsper}). }
\end{figure}

Figure \ref{ligo_sensitivity} gives the comparison
of  the sensitivity curve of the ground-based interferometer LIGO
with  the spectra of $\beta = -1.8$, $-1.9$, and $-2.02$
from  Fig.\ref{diff_beta}.
Here the vertical axis is the root mean square amplitude per root Hz,
which equals to
\be \label{rmsper}
\frac{h(\nu)}{\sqrt{\nu}}.
\ee
Note that, relevant to LIGO,
the frequency range is $(10 \sim  10^4)$ Hz,
which is not to be affected by the neutrino damping.
Obviously from the plot,
the LIGO I SRD  \cite{ligo} is yet not able to
detect the signals of RGWs in  the $\beta=-1.8$ model even
with a very large ratio $r=2.2$.
Therefore, LIGO is unlikely to able detect the RGWs,
as it currently stands.
The Advanced LIGO with greatly enhanced sensitivity \cite{ligo}
will be able to put a tighter constraints on the parameters.

\begin{figure}
\centerline{\includegraphics[width=8cm]{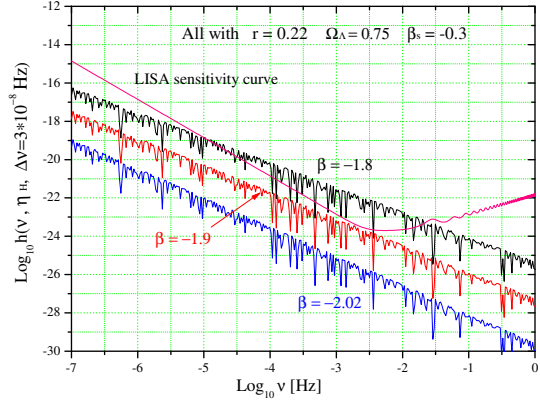}}
\caption{\label{lisa_sensitivity}
 Comparison of the spectra with
the LISA sensitivity curve~\cite{shane}.
The vertical axis is the r.m.s spectrum
defined in Eq.(\ref{rmssp}).
LISA will be able to detect the inflationary models
of $\beta=-1.8$ and $-1.9$. }
\end{figure}

Figure~\ref{lisa_sensitivity}
is a comparison of  the LISA  sensitivity curve
with the spectra from  Fig.\ref{diff_beta}
in the frequency range $(10^{-7}, 10^0)$ Hz.
Although these frequencies are lower than that for LIGO,
it is still not to be affected by the neutrino damping either.
Assuming that LISA has one year observation time,
which corresponds to frequency bin
$\Delta\nu=3\times 10^{-18}$Hz (i.e., one cycle/year)
around each frequency.
Thus, to make a comparison with the sensitity curve,
we need to rescale the spectrum $h(\nu)$ in Eq.(\ref{33})
into the root mean square spectrum $h(\nu, \Delta\nu)$
in the band $\Delta \nu$ \cite{grishchuk2000},
\be \label{rmssp}
h(\nu, \Delta\nu) = h(\nu)\sqrt{\frac{\Delta\nu}{\nu}}.
\ee
This r.m.s spectrum can be directly compared  with the 1 year
integration sensitivity curve that is downloaded from LISA \cite{shane}.
The plot shows that
LISA by its present design will be able to easily detect
the RGWs in the  inflationary model of $\beta=-1.8$.
If the ratio $r > 0.22$, LISA will also be able to
detect the inflationary models of $\beta = -1.9$.
However, LISA is unlikely to be able to detect the model
of $\beta=-2.02$.
Here Fig.~\ref{ligo_sensitivity} and Fig.~\ref{lisa_sensitivity} also
correct the mistake of Ref.\cite{zhang2},
where improper comparison is made
with the LIGO data and the LISA sensitive curve.
As will be seen in the following,
the neutrino free streaming practically affects only
the spectrum in a frequency range of $(10^{-16} \sim 10^{-10})$,
therefore, Fig.~\ref{ligo_sensitivity} and Fig.~\ref{lisa_sensitivity}
are not to be changed by neutrinos.

\begin{figure}
\centerline{\includegraphics[width=8cm]{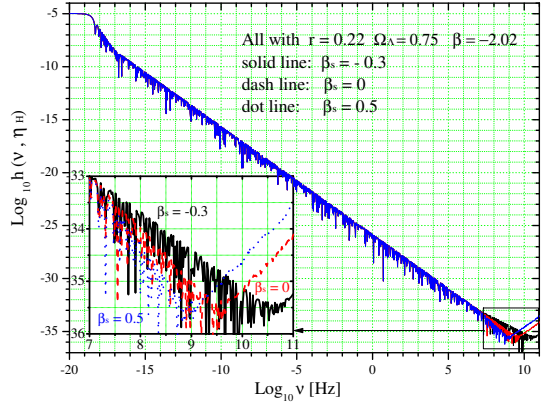}}
\caption{\label{diff_betas}
  The reheating affects the spectrum $h(\nu, \eta_H)$
only in very high frequency
range $\nu> 10^7$ Hz. }
\end{figure}

The influence of reheating stage on the spectrum
is shown in Fig.\ref{diff_betas}.
The spectra  for three different values of
$\beta_s=0.5$, $0$, $-0.3$ are given.
It is  clear that whereas the spectrum is almost unchanged
by $\beta_s$ in the large portion of
frequency range $\nu\leq 10^7$ Hz,
a larger $\beta_s$
will damp the amplitude in a high frequency range
$10^7\sim 10^9$Hz.
However, around $\nu \sim 10^9$Hz the spectrum begins
to increase  considerably.
This feature of RGWs in the GHz range
is very interesting,
as this high-frequency range of RGWs is the scientific gaol of
some  electromagnetic detecting systems,
such as the one using a Gaussian laser beam~\cite{fangyu},
or a circulating microwave beam \cite{cruise}.
However, the predicted spectrum for the very high frequency
range $\nu> 10^{10}$ Hz is not reliable,
since the energy scale of the conventional inflationary models
are less than $10 ^{16}$ Gev.

\begin{figure}
\centerline{\includegraphics[width=8cm]{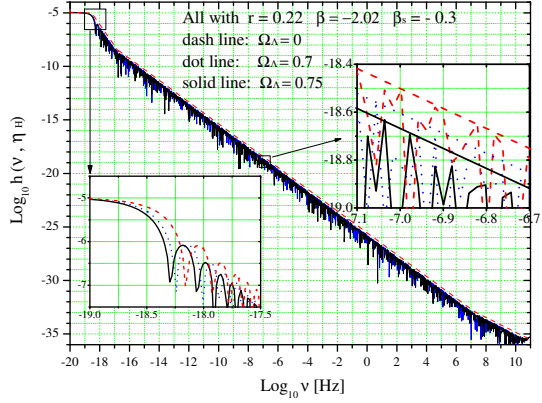}}
\caption{\label{diff_gamma}
The dependence of  $h(\nu, \eta_H)$
upon the dark energy $\Omega_\Lambda$ in the accelerating universe. }
\end{figure}

The influence of the dark energy on the spectrum $h(\nu,\eta_H)$
is demonstrated  in Fig.\ref{diff_gamma},
where $\Omega_{\Lambda}=0.0$, $0.7$, and $0.75$
are taken respectively.
Over the whole range of frequency $10^{-19} \sim 10^{10}$ Hz,
the amplitude of spectrum is altered by the presence of
$\Omega_\Lambda$, but the slope remains the same.
In regards to the amplitude,
firstly, the spectrum in a matter-dominant universe
of $\Omega_{\Lambda}=0$ is
higher than those in an accelerating universe
with $\Omega_{\Lambda}>0$, as Fig. \ref{diff_gamma} shows,
roughly by a factor $\sim 1.3$.
This feature is due to the fact the scale factor
$a(t_H)$  in  the accelerating  model is greater than
that in the matter-dominant models,
as has been explained at the ending paragraph of section 3.
Secondly, among the accelerating models,
by an analysis of the expression of $h(\nu,\eta_H)$,
the amplitude of $h(\nu,\eta_H)$
is proportional to $(\frac{a(\eta_E)}{a(\eta_H)})^3= 1/(1+z_E)^3$,
as has been explicitly shown in  Ref.\cite{zhang2}.
This phenomenon occurs basically because,
starting from the time $\eta_E$ up to the present time $\eta_H$,
the scale factor $a(\eta)$ increases by a different amount
in models of different $\Omega_\Lambda$ ;
thus, stretching of the physical wavelengths
and damping of the mode $h_k(\eta)$ are different correspondingly.
In the accelerating models with $\rho_\Lambda$ being constant,
one has approximately
$(\frac{a(\eta_E)}{a(\eta_H)})^3 \simeq \Omega_m/\Omega_\Lambda$;
thus,
the amplitude of $h(\nu,\eta_H) \propto \Omega_m/\Omega_{\Lambda}$,
i.e., the model with more dark energy component
has relatively a lower amplitude of $h(\nu,\eta_H)$.
Note that, in interpreting this relation
$h(\nu,\eta_H) \propto \Omega_m/\Omega_{\Lambda}$,
the dark energy $\Omega_\Lambda$ should be large enough,
say, $\Omega_\Lambda> \Omega_m $,
to ensure the sufficiently accelerating expansion.
This phenomenon is verified now in Fig.\ref{diff_gamma},
for example, the amplitude of
the $\Omega_{\Lambda}=0.75$ model
over that of the  $\Omega_{\Lambda}=0.7$ is found to be
$(\frac{0.25}{0.75})/ (\frac{0.3}{0.7})\approx0.8$.

\begin{figure}
\centerline{\includegraphics[width=8cm]{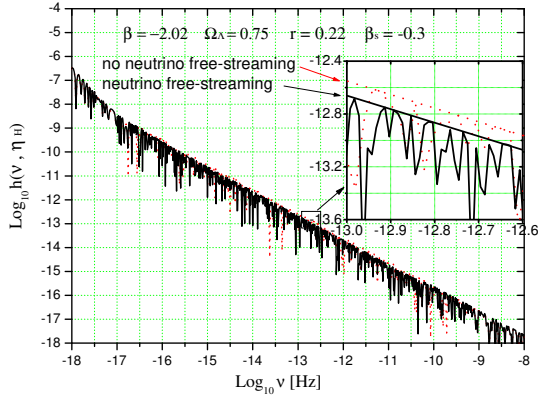}}
\caption{\label{neutrino_effect}
The neutrino free-streaming reduces
$h(\nu, \eta_H)$ in the frequency range $10^{-17}\sim 10^{-10}$ Hz. }
\end{figure}

Presented in Fig.\ref{neutrino_effect}
is the modification of the spectrum $h(\nu,\eta_H)$ by the
neutrino free-streaming up to the first order approximation
to Eq.(\ref{eq2}).
The effect is pronounced
in the low frequency range $(10^{-16}\sim 10^{-10})$ Hz,
where the amplitude of $h(\nu,\eta_H)$ is reduced
by a factor $\sim 20\%$ in comparison with the model
without neutrino free streaming.
Our analytical spectrum qualitatively agrees with
the numerical result in Ref.\cite{yuki} in the relevant range.
As we have mentioned earlier,
LIGO and LISA operating around $\sim 10^{2}$ Hz and $\sim 10^{-3}$ Hz,
respectively,
will not be able detect this neutrino damping.
But CMB anisotropies and polarization may be affected by that.

\begin{figure}
\centerline{\includegraphics[width=8cm]{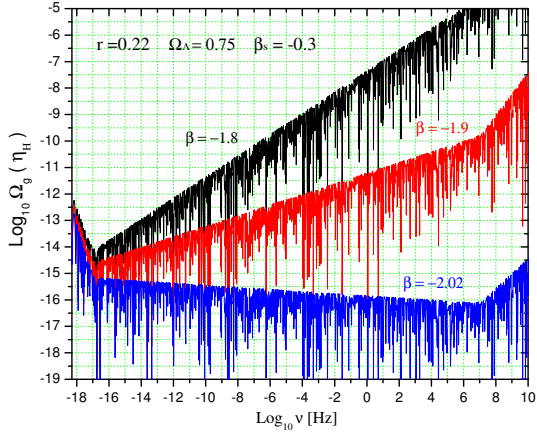}}
\caption{\label{diff_beta_Omega}
The spectral energy density $\Omega_g(\nu)$
for various inflation parameter $\beta$. }
\end{figure}

The dependence of spectral energy density
$\Omega_g$  in Eq.(\ref{32}) on the inflationary models is
illustrated in Fig.\ref{diff_beta_Omega}.
For the purpose of clarity,
the neutrino free streaming is not taken into account.
Clearly, $\Omega_g$ is very sensitive to
the parameter $\beta$,
and a larger $\beta$ gives a higher $\Omega_g$.
The model of $\beta=-1.8$ has an  $\Omega_g$ too high,
and as mentioned in paragraph before Eq.(\ref{BBN}),
it has  already been ruled out by the  nucleosynthesis
bound of Eq.(\ref{BBN}).
The Advanced LIGO~\cite{ligo}  will be able to detect RGWs
with $\Omega_g h^2>10^{-9}$ at $\nu \sim 100$Hz,
and it might impose stronger constraints on
other inflationary models.

\begin{figure}
\centerline{\includegraphics[width=8cm]{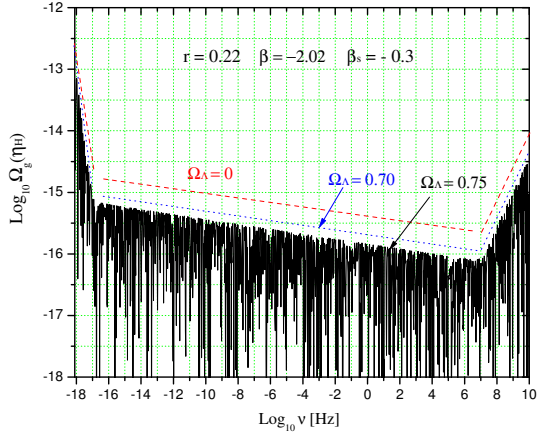}}
\caption{\label{diff_gamma_Omega}
The spectral energy density $\Omega_g(\nu)$
for different $\Omega_\Lambda$.
A larger $\Omega_\Lambda$
yields a lower $\Omega_g(\nu)$. }
\end{figure}

The impact of dark energy $\Omega_\Lambda$ on
the spectral energy density $\Omega_g$ is plotted in
Fig.\ref{diff_gamma_Omega} for the model $\beta=-2.02$.
It is clear seen that the accelerating expansion   of
the universe will cause a decrease of the amplitude of $\Omega_g$
over the whole range of frequencies,
and a larger $\Omega_\Lambda$ gives a lower $\Omega_g$.
Obviously, the effect due to the acceleration of expansion
of the  universe cannot be simply ignored.

\begin{figure}
\centerline{\includegraphics[width=8cm]{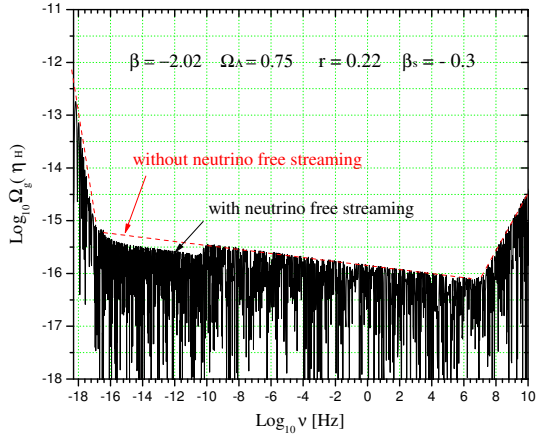}}
\caption{ \label{neutrino_effect_Omega}
The neutrino free streaming reduces
 $\Omega_g(\nu)$ in the range $10^{-17} \sim 10^{-10}$Hz.}
\end{figure}

The damping effect of neutrino free-streaming
on the spectral energy density $\Omega_g(\nu)$
is illustrated in Fig.\ref{neutrino_effect_Omega}
for the model $\beta=-2.02$.
The effect is mostly within the frequency
range $10^{-16} \sim 10^{-10}$Hz,
where the amplitude of $\Omega_g(\nu)$ drops visibly
by a factor of $\sim 36\%$.
Correspondingly,
the energy density of RGWs in Eq.(\ref{gwe}) is now reduces to
 $\Omega_{GW}=1.1\times 10^{-14}$ after considering
the neutrino free-streaming.
Recall that
it was $\Omega_{GW}=1.54\times 10^{-14}$ without neutrino damping,
thus the neutrino damping has caused a drop of $\sim 29 \%$ of
$\Omega_{GW}$.
Our result is also qualitatively consistent with the
numerical calculation in Ref.~\cite{yuki}.

ACKNOWLEDGMENT:
Y. Zhang's work has been supported by
the CNSF No.10173008,  NKBRSF G19990754, SRFDP, and CAS.
We thank Dr.L.Q. Wen at MPI for very helpful discussions.

\begin{center}
\textbf{Appendix}
\end{center}

In this appendix  we derive the anisotropic stress
tensor $\pi_{ij}$ of cosmic neutrinos during their free streaming,
filling in the details skipped in Ref.\cite{weinberg}.
Next we present the perturbation method to systematically
solve the equation of GRWs.
As a merit, this method applies for the realistic situation of
a time-dependent fractional
energy density $f_\nu(\eta)$ of neutrinos,
a nonzero decoupling time $u_{dec}\ne 0$,
and a non-vanishing time derivative $\chi'(u_{dec})\ne 0$.
This solution is an extension to the previous works
in Refs.\cite{weinberg}  \cite{dicus}.
The same notations as in Ref.\cite{weinberg} is used.

In the radiation stage at temperatures $\geq 2$ Mev ,
the neutrinos are
in equilibrium with other relativistic species,
such as electron and photons,
together forming the radiation component.
During this period,
practically all the $k$-modes $h_k(\eta)$ of cosmological interest
are still far outside the horizon,
thus remain a constant, $h_k(\eta)\simeq $ constant,
thus, are not effected by the neutrinos.
With the further expansion of the universe
when the temperature drops down to $<2$ Mev,
the neutrinos are going out of equilibrium
and are starting to stream freely in space.
Then the neutrinos will be able to influence
the modes $h_k(\eta)$ of short wavelengths
that re-enter the horizon.
The equation  of RGWs then becomes inhomogeneous
\be \label{hij}
h_{ij}''(\eta)+2\frac{a'(\eta)}{a(\eta)}h_{ij}'(\eta)
-\nabla^2 h_{ij}(\eta)
=16\pi Ga^2\pi_{ij}(\eta).
\ee
Here the source term $\pi_{ij}$, contributed by neutrinos,
 is the anisotropic part of the stress tensor $T_{ij}$
and is effective only during the period $\eta_{dec}\leq \eta\leq\eta_2$,
from the neutrinos decoupling up to
the beginning of the matter domination.
When the matter domination begins,
the neutrino number density  has been diluted out
by a factor $\sim 10^{- 3\times 6}$,
so the source $\pi_{ij}$ is effectively switched off
after the matter domination.
In terms of the neutrino distribution
function $n({\bf x, p}, t)$ and the momentum $p^i$,
the spatial part of the neutrino energy-momentum
stress tensor is written as
\be
 T^i_{\,\,j}= \frac{1}{\sqrt{-g}}
 \int d^3{\bf p}\,n({\bf x,p},t)p^ip_j/p^0.
\ee
To keep the same notation with Ref.\cite{weinberg},
here the cosmic time $t=\int a (\eta)d\eta$ is used.
In the presence of the perturbations $h_{ij}$ of the metric,
$n({\bf x,p},t)$, $p^i$, and $p^0$ all depends on $h_{ij}$.
So the  stress tensor is written as a sum of
\be
T^i_{j} = P\delta_{ij} +\pi_{ij},\,\,\,\,\, {i,j = 1,2,3}
\ee
where $P\delta_{ij}$ is the unperturbed part with
$P$ being the homogeneous and isotropic pressure,
and the anisotropic stress tensor $\pi_{ij}$ is
the perturbed part caused by $h_{ij}$
\be \label{piij}
\pi_{ij}({\bf x },t)=
\frac{1}{\sqrt{-g}}  \, \delta  \int d^3{\bf p}
\,n({\bf x,p},t)
\frac{p^ip_j}{p^0 }.
\ee
Since $h_{ij}$ is small,
only the first order of $h_{ij}$ is needed in evaluating $\pi_{ij}$.
The distribution function $n({\bf x,p},t)$
satisfies the Boltzmann equation
\be  \label{boltzmann}
\frac{\partial n}{\partial t}
+\frac{\partial n}{\partial x^i}\frac{dx^i}{dt}
+\frac{\partial n}{\partial p_i}\frac{dp_i}{dt}
=0.
\ee
With $dx^i/dt=  p^i/p^0$ and
the geodesic equation
$dp_{i}/d t= \frac{1}{2}g_{\gamma\delta,i}p^{\gamma}p^{\delta}
=\frac{1}{2}h_{jk,i}p^{j}p^k$,
Eq.(\ref{boltzmann})  can be expanded as
\be  \label{boltzmann2}
\frac{\partial n}{\partial t}
+\frac{p^i}{p^0}
\frac{\partial n}{\partial x^i}+
\frac{1}{2}h_{jk,i}\frac{p^jp^k}{p^0}
\frac{\partial n}{\partial p_i}
=0.
\ee
At the instant $t_{dec}$ of the  decoupling
the neutrinos are still an ideal gas,
so one writes
\be \label{n+}
n({\bf x,p},t)=n({\bf x,p},t_{dec})+
\delta n({\bf x,p},t),
\ee
where
\be
n({\bf x,p},t_{dec})=\frac{N}{(2\pi)^3}\Big[
\exp\Big(\sqrt{g^{ij}(\textbf{x},t_{dec})p_ip_j}/
T_{dec}\Big)+1\Big]^{-1}
\ee
is the distribution function of the ideal gas of temperature $T_{dec}$,
and $\delta n$ represents the perturbation
satisfying  $\delta n =0$ at $t_{dec}$.
In our treatment $p_i$ is treated as the unperturbed
momentum and $p^i=g^{ij}p_j$ as the perturbed one.
Substituting Eq.(\ref{n+}) into Eq.(\ref{boltzmann2}),
neglecting the higher order term
$\frac{1}{2}h_{jk,i}\frac{p^jp^k}{p^0}
\frac{\partial \delta n}{\partial p_i}$,
using $\partial n_{dec}/\partial t =0$,
\be \label{missed}
\frac{\partial n_{dec}}{\partial x^i}=-\frac{1}{2}\bar{n}'(p)\,p
\hat{p_j}\hat{p_k}
\frac{\partial}{\partial x^i}h_{jk}({\bf x},t_{dec}),
\ee
(Ref.\cite{weinberg} missed a factor $\frac{1}{2}$ in Eq.(\ref{missed})),
and $\partial n/\partial p_i=\bar{n}'(p)\hat{p_i}$,
where $\bar{n}'(p) \equiv \partial \bar{n}/\partial p$,
$\hat{p_i}\equiv p_i/p$ and $p\equiv\sqrt{p_ip_i}$,
the Boltzmann equation reduces to the following
\be \label{dn}
\frac{\partial\delta n}{\partial t}+\frac{\hat{p_i}}{a(t)}
\frac{\partial\delta n}{\partial x^i}=-\frac{p}{2a(t)}\bar{n}'(p)
\hat{p_i}\hat{p_j}\hat{p_k}\frac{\partial}{\partial x^k}
(h_{ij}({\bf x},t)-h_{ij}({\bf x},t_{dec})),
\ee
where
$\bar{n}(p)=\frac{N}{(2\pi)^3}
[\exp(p/T_{dec}a_{dec})+1]^{-1}$.
This equation can be decomposed into the Fourier $\bf k$-modes,
as in  Eq.(\ref{planwave}),
and each mode has
the formal solution
\be \label{formal}
\delta n_k({\bf p},\eta) =
 -\frac{i}{2}p\bar{n}'(p)\hat{p_i}\hat{p_j}
\hat{p}\cdot {\bf k}
\int^{\eta}_{\eta_{dec}}d\eta'
e^{i\hat{p}\cdot{\bf k}(\eta'-\eta)}
\Big[h_{ij}(k, \eta')-h_{ij}(k,\eta_{dec})\Big],
\ee
where the comoving time $d\eta=dt/a$ is used,
and the integrand function depends on $h_{ij}$ explicitly, where
\be
h_{ij}(k,\eta) \equiv
\sum_{\sigma}\epsilon^{\sigma}_{ij}h_k^{(\sigma)}(\eta).
\ee
Defining the variable $u\equiv k(\eta-\eta_{dec})$,
Eq.(\ref{formal})  just reduces  to Eq.(13)
in Ref.\cite{weinberg}.
There are four terms that contribute to $\pi_{ij}$
in Eq.(\ref{piij}).
Specifically,
the perturbations to the distribution function $n$ give two terms:
\be \label{67}
\frac{p^ip_j}{p^0}\delta n_k (\eta_{dec})
=-\frac{p^2}{2a}\bar{n}'(p)\hat{p_i}\hat{p_j}\hat{p_m}\hat{p_l}
h_{ml}(k, \eta_{dec}),
\ee
\be \label{68}
\frac{p^ip_j}{p^0}\delta n_k(\eta)
=-\frac{p^2}{2a}\bar{n}'(p)\hat{p_i}\hat{p_j}\hat{p_m} \hat{p_l}
\Big[ h_{ml}(k,\eta)-h_{ml}(k,\eta_{dec})
-\int^{\eta}_{\eta_{dec}}d\eta'e^{i\hat{p}\cdot\vec{k}(\eta'-\eta)}
h'_{ml}(k, \eta')\Big],
\ee
where integration by parts with respect to the time $\eta'$
has been used, the following two terms also contribute
\be \label{69}
\frac{\bar{n}(p)}{p^0}p_j\delta p^i
=  -\frac{p}{a}\bar{n}  \hat{p_l}\hat{p_j}  h_{il}(k,\eta) ,
\ee
\be \label{70}
-\frac{\bar{n}p^ip_j}{(p^0)^2}\delta p^0
=\frac{p}{2a}\bar{n}\hat{p_i}\hat{p_j}\hat{p_m}\hat{p_l}h_{ml}(k,\eta).
\ee
One puts these four terms from Eq.(\ref{67}) through
Eq.(\ref{70}) into Eq.(\ref{piij})
and carries out the integration  $\int d^3{\bf p }$.
The spherical coordinates with $z=  \bf \hat k$
can be used in doing the angular integration,
so that $\hat{p}\cdot\bf \hat{k}=\cos\theta\equiv\mu$.
Since $h_{ml}$ is transverse,
one has
\be
\int d\Omega\,\hat{p_i}\hat{p_j}\hat{p_m}\hat{p_l}h_{ml}
= \frac{\pi}{4}
(\delta_{im}\delta_{jl} +\delta_{il}\delta_{jm}) h_{ml}
\int^{1}_{-1}d\mu(1-\mu^2)^2.
\ee
After some calculation,  one arrives at the resulting
anisotropic stress
\be \label{pi}
\pi_{ij}=-4\bar{\rho}_\nu(\eta)
\int_{\eta_{dec}}^{\eta}d\eta'K(k\eta-k\eta')h'_{ij}(\eta'),
\ee
where $\bar{\rho}_\nu(\eta)=a^{-4}\int d^3p\,p\bar{n}(p)$
is the neutrino density,
and $K$ is the kernel defined as
\be \label{kernel}
K(s)=\frac{1}{16}\int_{-1}^1 d\mu \,(1-\mu^2)^2  e^{i\mu s}
  =\frac{j_2(s)}{s^2},
\ee
with $j_2(s)$ is the spherical Bessel function.
Since $h_{ij}$
is traceless and transverse, so is $\pi_{ij}$,  by Eq.(\ref{pi}),
with $\pi^i_{\,\,i}=0$ and $ \pi_{ij,j}=0$.
Substituting Eq.(\ref{pi})
into Eq.(\ref{hij}) and using the Friedmann equation
$a'\,^2/a^4=8\pi G\bar\rho/3$
yields the integro-differential
equation~\cite{weinberg}
\be \label{74}
h_k''(\eta)+\frac{2a'(\eta)}{a(\eta)}h_k'(\eta)+k^2h_k(\eta)
=-24f_\nu
(\eta)\Big[\frac{a'(\eta)}{a(\eta)}\Big]^2\int_{\eta_{dec}}^{\eta}
d\eta' K(k\eta-k\eta') h_k'(\eta'),
\ee
where the fractional neutrino energy density
\be \label{fraction}
f_\nu(\eta) \equiv \frac{\bar{\rho}_\nu(\eta)}{\bar{\rho}(\eta)}.
\ee
Although at present the dark energy $\Omega_\Lambda$ is dominant,
but it is negligible during the radiation-dominant stage,
in comparison with the matter, neutrino, and radiation components.
Even in the dynamic models of dark energy
evolving with time,
the contribution from $\rho_\Lambda(\eta)$
during radiation-dominant stage is not allowed to be
more than a few percent of
the total energy \cite{zhang-plb} \cite{amendola}.
Therefore,  Eq.(\ref{fraction}) is practically equal to
\be
 f_\nu(\eta) =\frac{f_\nu(0)}{1+a(\eta)/a_{eq}} \, ,
\ee
where $a_{eq}=a(\eta_2)$ is the scale factor
at the radiation-matter equality,
and
\be
f_\nu(0)= \frac{\Omega_\nu}{\Omega_\nu +\Omega_\gamma},
\ee
with  $\Omega_\nu$ and  $\Omega_\gamma$ being
the present fractional energy density o
f the neutrinos and the radiation, respectively.
Since $\eta_{dec}\gg\eta_e$, we can write $a(\eta)=a_e\eta$
with $a_e$ defined in Eq.(\ref{r}) for the radiation-dominant stage.
Introducing the variable $u\equiv k\eta$ and setting
\be h_k(u) =h_k(\eta_{dec})\chi(u),
\ee
then Eq.(\ref{74}) is reduced to
\be\label{77}
\chi''(u)+\frac{2}{u}\chi'(u)+\chi(u)
=-24 \frac{f_\nu(0)}{u^2(1+\alpha\,u)} \int_{u_{dec}}^{u}d U
K(u-U)\chi'(U),
\ee
where $\alpha\equiv a_e/k\,a_{eq}$.

Dicus \& Repko \cite{dicus} present an analytical solution
of Eq.(\ref{77}) under the approximation of setting
 $u_{dec}=0$, $\alpha=0$ and $\chi'(\eta_{dec})=0$.
This is only  valid for those modes that reenter
the horizon well after the neutrino decoupling.
Note that, in the coordinate that is used,
the decoupling time $u_{dec} \neq 0$ as in Eq.(\ref{aling1}).
Besides, for the short wavelength modes the derivative
$\chi'(\eta_{dec}) \neq 0$.
Moreover, actually $\alpha u=1$ at the time $\eta_2$,
thus, setting $\alpha = 0$ for
the whole period $( \eta_{dec} , \eta_{2})$
would lead to an over-account of
the fractional neutrino energy density $f_\nu(\eta)$
and consequently would give a lower amplitude of RGWs.
Unlike in Refs.\cite{weinberg,dicus} ,
here  we do not make the above-mentioned approximation,
 but instead,
keep $u_{dec}$, $\alpha $ and $\chi'(\eta_{dec})$ as they are.
We use a perturbation method to solve the integro-differential equation
(\ref{77}) analytically,
which can achieve high accuracy as one requires.
Note that the source term on the r.h.s. of
Eq.(\ref{77}) is relatively small,
and setting it to be zero yields  the homogeneous equation
\be
\chi_0''(u)+\frac{2}{u}\chi_0'(u)+\chi_0(u)=0,
\ee
with the solution
\be \label{homo}
\chi_0(u)=c_1\frac{e^{iu}}{u}+c_2\frac{e^{-iu}}{u},
\ee
as the $0$$^{th}$ order approximation to Eq.(\ref{77}),
where $c_1$ and $c_2$ are the coefficients determined
by the continuity condition at $u_{dec}$.
One substitutes $\chi'_0(u)$ in place of $\chi'(u)$
in the integration of Eq.(\ref{77})
to give the $1^{st}$ order approximation
\be \label{82}
\chi_1''(u)+\frac{2}{u}\chi_1'(u)+\chi_1(u)=-24
\frac{f_\nu (0)}{u^2(1+\alpha\,u)}
\int_{u_{dec}}^{u}d U K(u-U)\chi_0'(U),
\ee
which is a differential equation with a known inhomogeneous term.
It has a particular solution
\begin{eqnarray} \label{particular}
\chi^*(u)&=&\int^u_{u_{dec}}
\frac{y_2(u)y_1(v)-y_1(u)y_2(v)}{W[y_1,y_2](v)}r(v)dv\nonumber\\
&=&-\frac{24f_{\nu}(0)}{u}
\int^u_{u_{dec}}dv\frac{\sin(u-v)}
{v(1+\alpha v)}\int^{v}_{u_{dec}}ds
\frac{j_2(v-s)}{(v-s)^2}\chi_0'(s),
\end{eqnarray}
where $r(v)$ represents the inhomogeneous term of Eq.(\ref{82}),
and $y_1=\frac{e^{i u}}{u}$, $y_2=\frac{e^{-i u}}{u}$
are the two linearly independent
solutions to the homogeneous counterpart,
and $W[y_1,y_2](v)=-2i/v^2$ is the Wronskian.
Therefore, the solution of Eq.(\ref{82}) is given by
\be \label{first}
\chi_1(u)=\chi_0(u)+\chi^*(u),
\ee
which is also the $1^{st}$ order approximate solution of Eq.(\ref{77}).
Similarly, one substitutes the $1^{st}$ order solution
$\chi_1(u)$ into Eq.(\ref{particular}),
and obtains  the $2^{nd}$ order approximate equation,
\be \label{85}
\chi_2''(u)+\frac{2}{u}\chi_2'(u)+\chi_2(u)=-24
\frac{f_\nu (0)}{u^2(1+\alpha\,u)}
\int_{u_{dec}}^{u}d U K(u-U)\chi_1'(U),
\ee
which has a particular solution
\be \label{particular2}
\chi^{**}(u) =
-\frac{24f_{\nu}(\eta_{dec})}{u}\int^u_{u_{dec}}
dv\frac{\sin(u-v)}
{v(1+\alpha v)}\int^{v}_{u_{dec}}ds
\frac{j_2(v-s)}{(v-s)^2}\chi_1'(s),
\ee
and thus the  $2^{nd}$ order approximate solution is
\be \label{second}
\chi_2(u)=\chi_0(u)+\chi^{**}(u).
\ee
By the same routine,  the higher order solutions can be obtained.
In fact, as our calculation shows that
the $1^{st}$ order approximation is already accurate enough
for the purpose of computing the spectrum for RGWs.
An important advantage of our solution is that
it is valid for those modes that reenter the horizon before or
after the decoupling time $\eta_{dec}$.
Integrations, such as  in Eqs.(\ref{particular}) and (\ref{particular2}),
can be done easily by common computing tools.

\end{document}